\documentclass[%
reprint,
superscriptaddress,
 amsmath,amssymb,
 aps,
pra,
]{revtex4-1}
\usepackage{graphicx}
\usepackage{dcolumn}
\usepackage{bm}
\usepackage{color} 
\newcommand{\beq}{\begin{equation}}
\newcommand{\eeq}{\end{equation}}
\newcommand{\beqa}{\begin{eqnarray}}
\newcommand{\eeqa}{\end{eqnarray}}

\begin{document}

\preprint{APS/123-QED}

\title{Electronic beam shifts in monolayer graphene superlattice}

\author{Xi Chen}
\email[]{xchen@shu.edu.cn}
\affiliation{Department of Physics, Shanghai University,
200444 Shanghai, P. R. China}

\affiliation{Departamento de Qu\'{\i}mica F\'{\i}sica, UPV-EHU, Apdo 644, E-48080 Bilbao, Spain}

\author{Pei-Liang Zhao}%

\affiliation{Department of Physics, Shanghai University,
200444 Shanghai, P. R. China}

\author{Xiao-Jing Lu}%

\affiliation{Department of Physics, Shanghai University,
200444 Shanghai, P. R. China}

\author{Li-Gang Wang}

\affiliation{Department of Physics, Zhejiang University, Hangzhou, 310027, P. R. China}

\date{\today}

\begin{abstract}
Electronic analogue of generalized Goos-H\"{a}nchen shifts is investigated in the monolayer graphene superlattice with one-dimensional
periodic potentials of square barriers. It is found that the lateral shifts for the electron beam transmitted through the
monolayer graphene superlattice can be negative as well as positive near the band edges of zero-$\bar{k}$ gap,
which are different from those near the band edges of Bragg gap. These negative and positive beam shifts have close relation to the Dirac point.
When the condition $q_A d_A= -q_B d_B= m \pi$ ($m=1,2,3...$) is satisfied,
the beam shifts can be controlled from negative to positive when the incident energy is above the Dirac point, and vice versa.
In addition, the beam shifts can be greatly enhanced by the defect mode inside the zero-$\bar{k}$ gap. These intriguing phenomena
can be verified in a relatively simple optical setup, and have potential applications in the graphene-based electron wave devices.

\end{abstract}



\maketitle


\section{Introduction}

Since the first synthesis of graphene in 2004 \cite{Novoselov,Zhang-TS,Castro},
electronic analogies of optical phenomena \cite{Cheianov,Park,Garcia,Shytov,Darancet,Ghosh,Dragoman,Zhao,Chen-APL,Wu-APL,Williams},
such as reflection, refraction, polarization, interference, waveguiding,
focusing and collimation have inspired many interesting and new concepts on ``Dirac electron wave optics" to design the
graphene-based nanoelectronic devices, due to the link between
the Klein tunneling \cite{Katsnelson-NG} and negative refraction \cite{Guney}.
The close relation between electron and light results from the quantum-mechanical wave nature of electrons.
Among all these optical-like phenomena, Goos-H\"{a}nchen shift, referring to
the lateral shift for totally reflected light beam at a single dielectric interface \cite{Goos},
has been extended to the field of electronics \cite{Chen-PRB} and currently investigated in various graphene nanostructures \cite{Beenakker-PRL,Manish,Chen-EPJB,Wu,Zhai}.
The electronic lateral shifts result in a remarkable $8e^2/\hbar$ conductance plateau in $p$-$n$-$p$ graphene junction \cite{Beenakker-PRL},
and further provide an alternative way to realize valley beam splitter in strained graphene \cite{Wu,Zhai}.

Motivated by the experimental progress on the construction of graphene superlattice, 
the electronic bandgap and transport in graphene superlattice have attracted much attention \cite{Bai,Barbier2008,Barbier2009,Brey,Park2008,Park2009,Bliokh,Wang2010}.
Interestingly, the graphene superlattice is analogous to one-dimensional photonic crystal containing positive-index and negative-index materials \cite{Bliokh,Wang2010},
thus there exists zero-averaged wave number (zero-$\bar{k}$) bandgap in graphene superlattice. Like the zero-$\bar{n}$ gap in the photonic crystals with negative-index materials \cite{Li},
this so-called zero-$\bar{k}$ gap, which is associated with the Dirac point in graphene superlattice \cite{Barbier2009,Wang2010}, is different from Bragg gap and
thus robust against the lattice parameters and incidence angles. Besides the unique electronic band gap and transport in the graphene superlattice, the beam supercollimatation is
also achieved in two-dimensional graphene superlattice, utilizing the highly developed concepts in optics \cite{Park}.

In this work, we will investigate the generalized Goos-H\"{a}nchen shifts of electron beam in transmission through monolayer graphene superlattice with one-dimensional
periodic potentials of square barriers. The beam shifts discussed here has nothing to do with the evanescent wave and total reflection, so we term it as generalized Goos-H\"{a}nchen shift,
since the lateral shift, resulting from the finite width of the beam, keeps the main features of the Goos-H\"{a}nchen shift.
The behaviors of beam shifts are quite different near the band edges of the zero-$\bar{k}$ gap and Bragg gap.
What we emphasize here is that the unusual feature of beam shifts around the Dirac point in the graphene superlattice,
that is to say, when the incident energy is below the Dirac point,
the beam shifts can be positive and vice versa. Moreover, there will exists transition between the negative and positive beam shifts, depending on
the condition $q_A d_A= -q_B d_B= m \pi$ ($m=1,2,3...$). When the defect in the graphene superlattice is further induced,
the beam shifts inside the zero-$\bar{k}$ bandgap can be greatly enhanced near the defect mode, whose location depends on the incidence
angles at a fixed incident energy. Furthermore, the beam shifts can also be controlled by adjusting the
potential height of defect in the graphene superlattice. Consequently, the tunable negative and positive beam shifts will provide the potential applications
in graphene-based electron wave devices (e.g., beam switch, beam splitter and spatial modulator).

\section{Theoretical  Model}

\begin{center}
\begin{figure}[]
\scalebox{0.55}[0.55]{\includegraphics{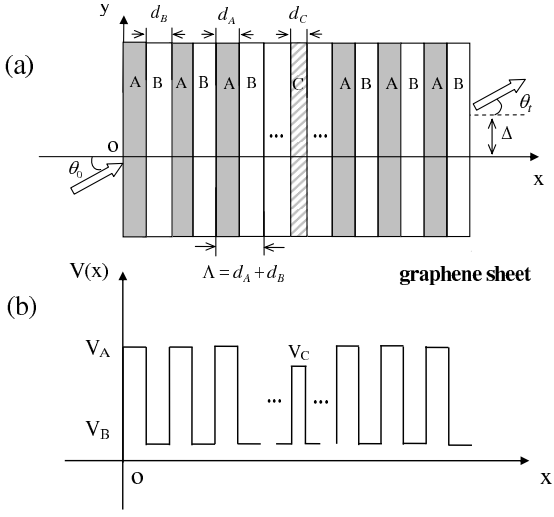}}
\caption{(Color
online) (a) Schematic diagram for beam shift of the transmitted electrons through a graphene superlattice with defect $C$.
(b) The profiles of the potentials.
\label{fig.1}}
\end{figure}
\end{center}

Consider a graphene-based superlattice with one-dimensional periodic squared potential
barriers. The schematic structure for the monolayer
graphene superlattice is shown in Fig. \ref{fig.1}, where the parameters are
the potential barrier height $V_{A}$ and width $d_A$, the potential well height $V_B$ and width $d_B$, and
the defect is denoted by the potential height $V_{C}$ and width $d_{C}$. In what follows,
we will first study the lateral shift in the graphene superlattice $(AB)^{N}$, and discuss further
the influence of defect mode in the structure of $(AB)^{N}C(BA)^{N}$.

In general, the Hamiltonian of electrons inside a monolayer graphene, in the vicinity of the $K$ point
and in the presence of a potential $V(x)$, is given by the following massless Dirac-like equation,
\beq
\hat{H}=-i\hbar v_{F} {\vec{\sigma} \cdot \vec{\nabla} }+V(x),
\eeq
where the Fermi velocity $v_{F}\approx 10^{6}$m/s, and $\vec{\sigma}%
=(\sigma _{x},\sigma _{y})$ are the Pauli matrices. Due to the translation invariance in the $y$ direction,
the solution of above equation
for a given incident energy $E$ and potential barrier $V_j$ can be presented as $\tilde{\Psi} (x,y) = \Psi (x)  e^{i k_{y}y}$ with
\begin{equation}
\label{barrier}
\Psi (x) = \left[
 a_i e^{i q_{j}x} \left(\begin{array}{c}
      1 \\
      \frac{q_j +ik_y}{k_j} \\
   \end{array}\right)
+
b_i e^{-iq_{j}x}
 \left(\begin{array}{c}
      1 \\
      \frac{-q_j +ik_y}{k_j} \\
   \end{array}\right)
 \right],
\end{equation}
where the spinor wave function $\Psi=(\Psi_{+}, \Psi_{-})$ has the pseudospin component $\Psi_{+}$ and $\Psi_{-}$ on the $A$ and $B$ sublattice,
$k_j=(E-V_j)/\hbar v_F$, $k_{y}$ and $q_j$ are the $y$ and $x$ components of wave number,
$q_{j}= \mbox{sign}(k_j) ({k^2_{j}-k_{y}^{2}})^{1/2}$ for $ k^2_{j}> k^2_{y}$,
otherwise $q_{j}=i  (k^2_{y}-k_{j}^{2})^{1/2}$, and $a_j$ ($b_j$) is the amplitude of the forward (backward)
propagating wave. Noting that when $E<V_j$, the wave vector is opposite to the direction of electron's velocity.
This results in the negative refraction in $p$-$n$ junction, and Veselago lens \cite{Cheianov}, which is relevant to the negative
Goos-H\"{a}nchen shifts in graphene barrier \cite{Chen-EPJB}.
The wave functions at any two positions $x$ and $x +\Delta x $ inside the $j$th potential can
be related by the transfer matrix \cite{Wang2010}:
\begin{equation}
\label{barrier}
M_j =  \left(\begin{array}{cc}
      \frac{\cos(q_j \Delta x - \theta_j)}{\cos \theta_j} & i \frac{\sin(q_j \Delta x)}{\cos \theta_j}
      \\
      i \frac{\sin(q_j \Delta x)}{\cos \theta_j} & \frac{\cos(q_j \Delta x + \theta_j)}{\cos \theta_j}
   \end{array}\right),
\end{equation}
with $\theta_j = \arcsin(k_y/k_j)$. After lengthy but direct calculations, the reflection and transmission coefficients are found to be
\beqa
r (k_y) &=&\frac{(\chi_{22} e^{- i\theta_0}-\chi_{11} e^{i\theta_t})-\chi_{12} e^{ i(\theta_t-\theta_0)}+\chi_{21}}{(\chi_{22} e^{- i\theta_0}+\chi_{11} e^{i\theta_t})-\chi_{12} e^{ i(\theta_t-\theta_0)}-\chi_{21}},
\\
t (k_y) &=& \frac{2 \cos \theta_0}{(\chi_{22} e^{- i\theta_0}+\chi_{11} e^{i\theta_t})-\chi_{12} e^{ i(\theta_t-\theta_0)}-\chi_{21}},
\eeqa
where $\theta_0$ and $\theta_t$ are incidence and exit angles (see Fig. \ref{fig.1}), and $\chi_{ij} (i,j=1,2)$ is the matrix element of total transfer matrix, $X_{N}= \prod^{N}_{j=1} M_j$, connecting the incident and exit ends,
and $N$ is the total number of the layers of graphene superlattice.

To calculate the generalized Goos-H\"{a}nchen shift, we consider an electron beam,
\begin{equation}
\label{incident beam}
\Psi^{in} = \frac{1}{\sqrt{2 \pi}}\int^{\infty}_{-\infty} d k_y A(k_y) e^{i (k_x x+k_y y)} \left(\begin{array}{c}
         1 \\
      e^{i \theta} \\
   \end{array}\right),
\end{equation}
where $k_x$ and $k_y$ are the $x$ and $y$ components of wave number in the incident region,
$k_x= (k^2- k^2_y)^{1/2}$, $\theta = \arcsin{(k_y/k)}$ and $k=E/\hbar v_F$ with assuming $V=0$. The angular-spectrum distribution is
assumed to be, but not necessarily, $A(k_y)=w_y \exp{[-(w^2_y/2)(k_y-k_{y0})^2]}$ around $k_{y0}$ corresponding to the incidence angle $\theta_0$,
where $w_y= w \sec \theta_0$, and $w$ is the half beam width at waist.
When the electron beam is well-collimated, there will be a narrow distribution of $k_y$ values around $k_{y0}$. Thus, we can expand
$k_x$ and $\theta$ to first order around $k_{y0}$ and substitute into Eq. (\ref{incident beam}). By evaluating the Gaussian integral,
we finally obtain the spatial profile of the incident beam at $x=0$, which leads to the two components as follows,
$
\Psi^{in}_{\pm} = \exp{[- (y-\bar{y}^{in}_{\pm})^2/ 2 w^2_y]},
$
with $\bar{y}^{in}_{+}=0$ and $\bar{y}^{in}_{-}= -\theta'(k_{y0})$. Noting that $\theta'(k_{y0})$
means the values of derivative with respect to $k_y$ taken at $k_y=k_{y0}$. Obviously, the separation of the two centers
is $\delta_0= |\bar{y}^{in}_{+} -\bar{y}^{in}_{-} |= -\theta'(k_{y0})$.
For the convenience, we can consider the potential in the transmitted region is the same as that in the incident one, so that the
transmitted beam can be then expressed as
\begin{equation}
\label{transmitted beam}
\Psi^{tr} =  \frac{1}{\sqrt{2 \pi}} \int^{\infty}_{-\infty} d k_y t(k_y) A(k_y)  e^{i [k_x (x-d_T)+k_y y]} \left(\begin{array}{c}
         1 \\
      e^{i\theta} \\
   \end{array}\right),
\end{equation}
where $d_T$ is the total length of the graphene superlattice and the complex transmission coefficient $t(k_y)$ for the relevant plane wave component can be expressed
in term of amplitude $\rho(k_y)$ and phase $\varphi(k_y)$, $t(k_y)= \rho(k_y) \exp{[i \varphi(k_y)]}$. As a matter of fact, when the
stationary phase approximation is applied, it is assumed that the phase $\varphi(k_y)$ of the transmission
coefficient $t(k_y)$ is linearly dependent on $k_y$ and the amplitude of the transmission keeps almost constant,
thus it is reasonable to expand the $\varphi(k_y)$ in Taylor series at $k_{y0}$, and retain up to the first-order
term. Similarly, $k_x$ and $\theta$ can be expanded to first order around $k_{y0}$. After substituting all into Eq. (\ref{transmitted beam}),
we can calculate the Gaussian integral, and obtain the two components of the transmitted beam at $x=d_T$,
$
\Psi^{tr}_{\pm} = \exp{[- (y-\bar{y}^{tr}_{\pm})^2/ 2 w^2_y]},
$
with $\bar{y}^{tr}_{+} = -\varphi'(k_{y0}) $ and $\bar{y}^{tr}_{-} = -\varphi'(k_{y0})-\theta'(k_{y0})$. Noting that the separation of the two centers
is also $\delta_0$, which is the same as that for incident beam.

Comparison of the incident and transmitted beams,
the displacement of two components are the same, $\sigma_{\pm} = |\bar{y}^{tr}_{\pm}-\bar{y}^{in}_{\pm}|= -\varphi'(k_{y0})$, so that the lateral shift of the electron beam transmitted through the graphene superlattice can be defined as \cite{Beenakker-PRL}, $\Delta= (\sigma^{+}+\sigma^{-})/2$, and thus is given by \cite{Chen-PRB,Chen-EPJB}
\begin{equation}
\label{lateral shift}
\Delta = - \varphi'(k_{y0}) = - \mbox{Im} \left( \frac{\partial}{\partial k_{y0}} \ln t \right),
\end{equation}
where $\partial/\partial k_{y0}$ denotes the derivative with respect to
$k_y$ evaluated at $k_y = k_{y0}$.

\begin{figure}[]
\begin{center}
\scalebox{0.7}[0.7]{\includegraphics{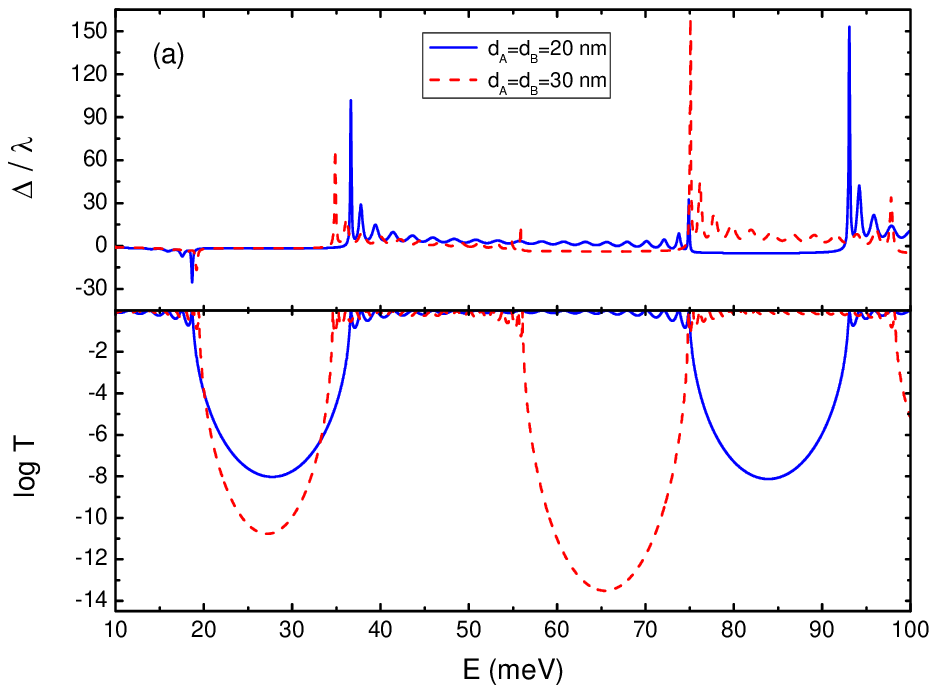}}
\scalebox{0.40}[0.40]{\includegraphics{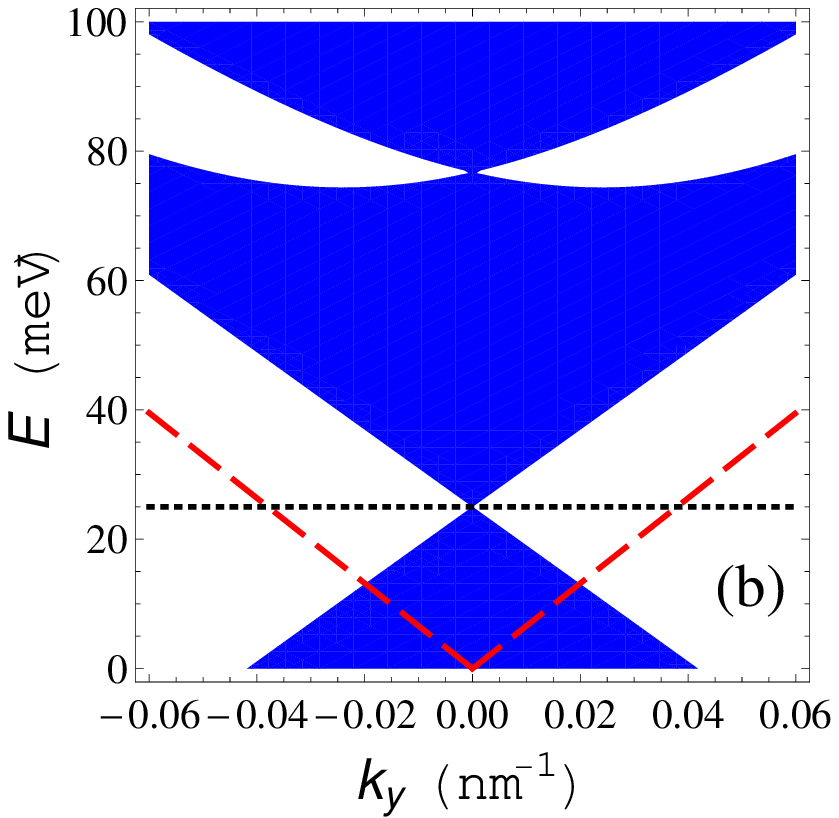}}
\scalebox{0.40}[0.40]{\includegraphics{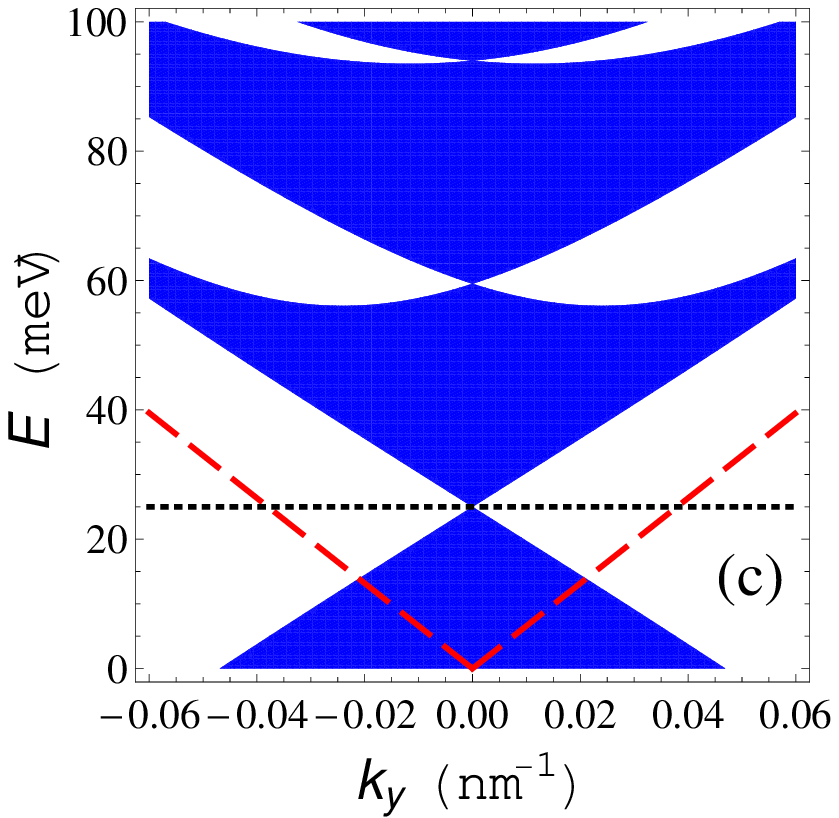}}
\caption{(Color online) (a) Dependence of beam shifts and transmission probability for transmitted electron beams in graphene superlattice $(AB)^{20}$ on the incident energy $E$,
where $\theta_0=20^{\circ}$, $V_A =50$ meV, $V_B=0$ meV, $d_B/d_A=1$, $d_A=20$ nm (solid blue line), and $d_A=30$ nm (dashed red line);
(b) and (c) are the corresponding electronic band structures for $d_A=20$ nm and $d_A=30$ nm.
The dot lines denote the location of the Dirac point $E_D=25$ meV, and the dashed lines present the ``light cones" of the incident electrons.}
\label{fig.2}
\end{center}
\end{figure}

\section{Results and Discussions}

Fig. \ref{fig.2} (a) shows the typical beam shifts in transmission through the graphene
superlattice $(AB)^{20}$, where $\theta_0=20^{\circ}$, $V_A =50$ meV and $V_B=0$ meV, $d_B/d_A=1$, $d_A=20$ nm (solid blue line), and
Fig. \ref{fig.2} (b) and (c) illustrates the corresponding electronic band-crossing structures in which $E_D = 25$ meV.
Surprisingly, the beam shifts near the band edges of the zero-$\bar{k}$ gap and Bragg gap are quite different. As shown in Fig. \ref{fig.2}, the beam shifts can be negative as well as positive near the band edges of the zero-$\bar{k}$ gap, which depends on the incident energy below or above the Dirac point, $E_D$. Whereas, the beam shifts near the band edges of the Bragg gap are positive. Remarkably, the enhanced beam shifts in this case are sensitive to
the lattice constants, $d_A$ and $d_B$, because the Bragg gap depends strongly on the lattice constants.

To understand beam shifts in such a graphene superlattice in term of electric band gap,
we write down the electronic dispersion at any incidence angles for the infinite superlattice $N \rightarrow \infty$, based on the Bloch's theorem,
in the following form:
\beqa
\label{dispersion}
\nonumber
\cos{(\beta_x \Lambda)} &=& \cos{(q_{A}d_{A}+q_{B}d_{B})} + \frac{\cos{(\theta_{A}-\theta_{B})}-1}{\cos{\theta_A} \cos{\theta_B}}
\\ &&\times \sin{(q_{A}d_{A})}\sin{(q_{B}d_{B})},
\eeqa
where $\beta_x$ is the $x$ component of Bloch wave vector, and $\Lambda=d_A +d_B$ is the length of the unit cell.
When $q_A d_A = -q_B d_B \neq m \pi$ ($m=1,2,3...$), and $\theta_A  \neq 0$, there is no real solution for $\beta_x$, thus there exists zero-$\bar{k}$ bandgap \cite{Wang2010}.
The Dirac point, the location of the touching point of the band, is given by  $q_A d_A= - q_B d_B$ at $\theta_A=0$.
In the simple case of $V_A \neq 0$ and $V_B =0$,  the Dirac point is located at $E_D = V_A/(1+d_B/d_A)$. For instance, the Dirac point is calculated as
$E_D=25$ meV with the parameters $V_A=50$ meV and $d_B/d_A=1$. From the electronic band structures in Fig. \ref{fig.2} (b) and (c), one can predict
that the group velocity is always negative (positive), when the incident energy is below (above) the Dirac point, based on the electronic dispersion relation.
So this provides the intuitive explanation of the negative and positive beam shifts in this case.

\begin{figure}[]
\begin{center}
\scalebox{0.70}[0.70]{\includegraphics{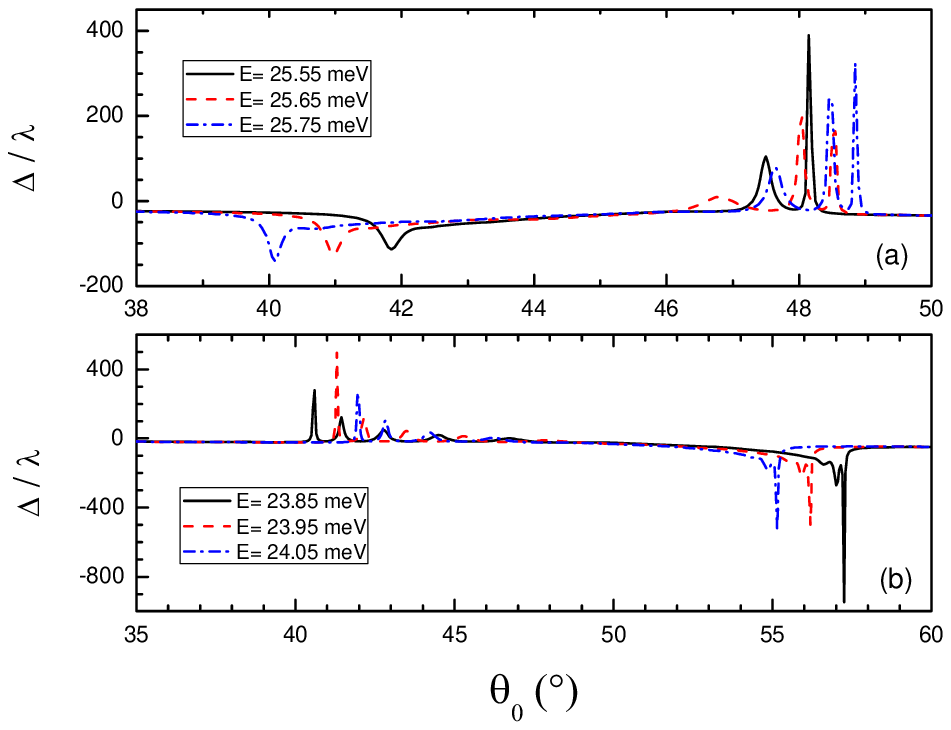}}
\scalebox{0.55}[0.52]{\includegraphics{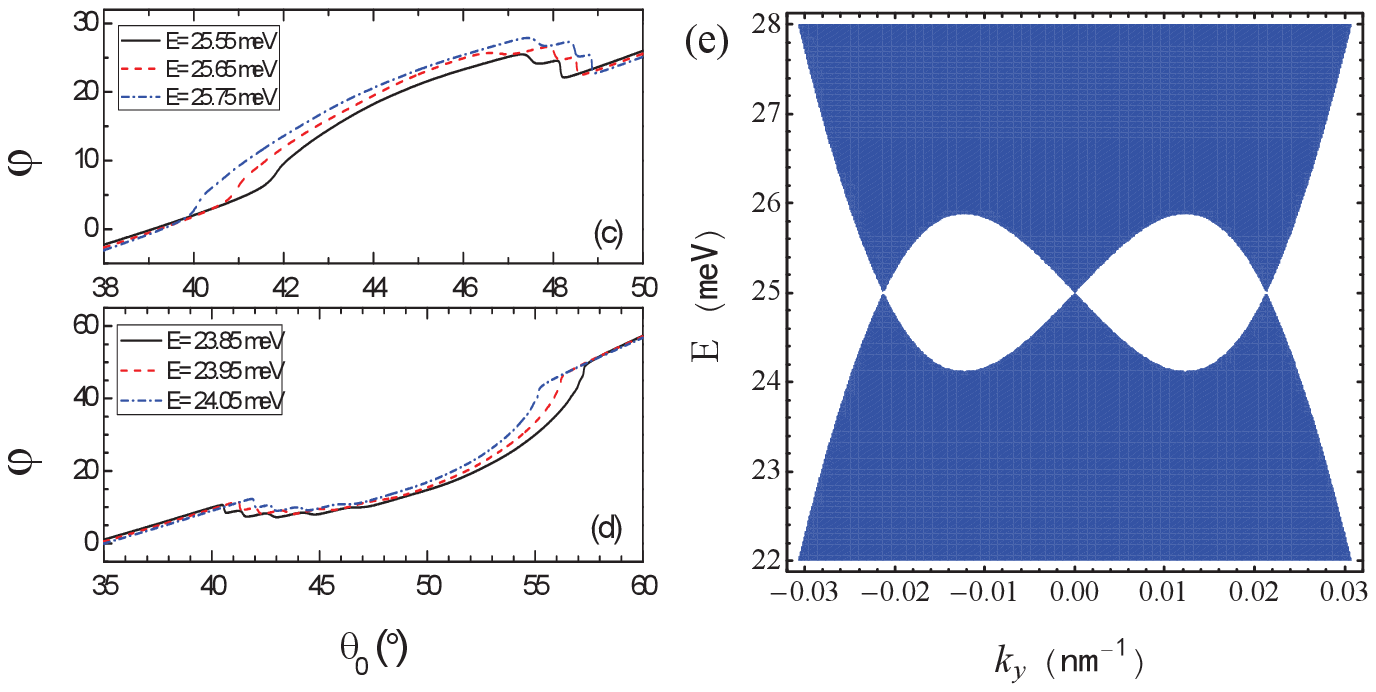}}
\caption{(Color online) Dependence of the beam shifts on the incidence angle near the extra Dirac point, where
(a) $E>E_D$ and (b) $E<E_D$. (c) and (d) are the corresponding phase shifts. (e) Electronic band structure for the graphene superlattice
with $d_B=d_A=120$ nm.}
\label{fig.3}
\end{center}
\end{figure}

More interestingly, according to the electronic dispersion (\ref{dispersion}), when the condition
\beq
\label{condition}
q_A d_A = - q_B d_B = m \pi, ~(m=1,2,3...),
\eeq
is satisfied, the zero-$\bar{k}$ gap will be closed, and a pair of new zero-$\bar{k}$ states emerges from $k_y=0$.
Actually, the extra Dirac points, located at the incline angles, for instance $\pm k_y \neq 0$, are only dependent on
the lattice constant $\Lambda=d_A+d_B$ and the ratio of $d_B/d_A$ \cite{Wang2010}. This condition (\ref{condition}) is the
same as that in one-dimensional photonic crystals containing negative-index and positive-index materials \cite{Wang-Zhu}.
Comparison of the electronic band structures for different lattice constants demonstrates
that the zero-$\bar{k}$ gap can be close and open periodically with the increasing the lattice constants. In the following discussion,
we shall focus on the beam shifts near the band-crossing structure, when the graphene superlattice satisfies the
condition (\ref{condition}), which offers the freedom to
control the lateral shifts around the Dirac points at non-normal incidence angles.

Fig. \ref{fig.3} (a) illustrates the dependence of the beam shifts on the incidence angle in the graphene superlattice, when the condition (\ref{condition}) with
$m=1$. Above the Dirac point ($E>E_D$), there exists transition from the negative beam shifts to the positive ones
with increasing the incidence angle $\theta_0$. On the contrary, the beam shifts changes from positive to negative below the Dirac point ($E<E_D$).
To confirm these intriguing phenomena, the corresponding phase shifts of the transmitted beam are depicted in Fig. \ref{fig.3} (c) and (d),
in which the slopes of the phase shifts suggest the negative and positive lateral shifts. As a matter of fact, the beam shift can be
understood from the reshape of the transmitted beam whose plane components undergo the different phase shifts.
These beam shifts are quite different from the previous ones, for example, $d_A=d_B=20$ nm. In that case, $q_A d_A =-q_B d_B < \pi$, so
the beam shifts near the band edges of the zero-$\bar{k}$ gap are always negative and positive, depending on whether the incident energy is above or below the Dirac point.
However, there is extra Dirac point at $k_y \neq 0$, as shown in Fig. \ref{fig.3} (e). Due to the periodical appearance of the Dirac
point, the group velocity can be negative and positive with increasing the incidence angle. As a consequence, the beam shifts
can be controlled from negative to positive, and vice versa.

\begin{figure}[]
\begin{center}
\scalebox{0.65}[0.65]{\includegraphics{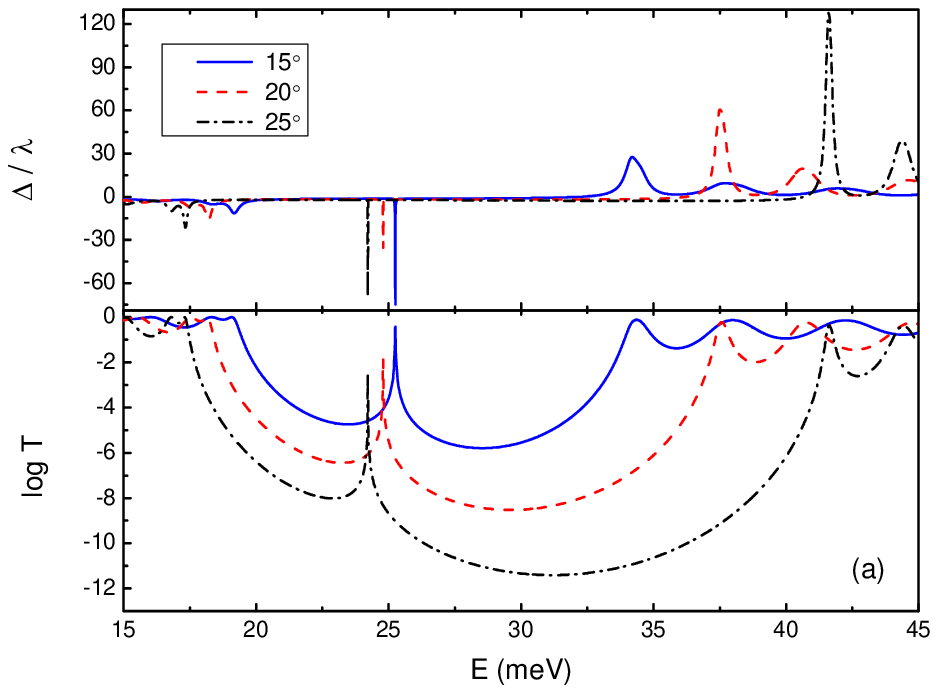}}
\scalebox{0.65}[0.65]{\includegraphics{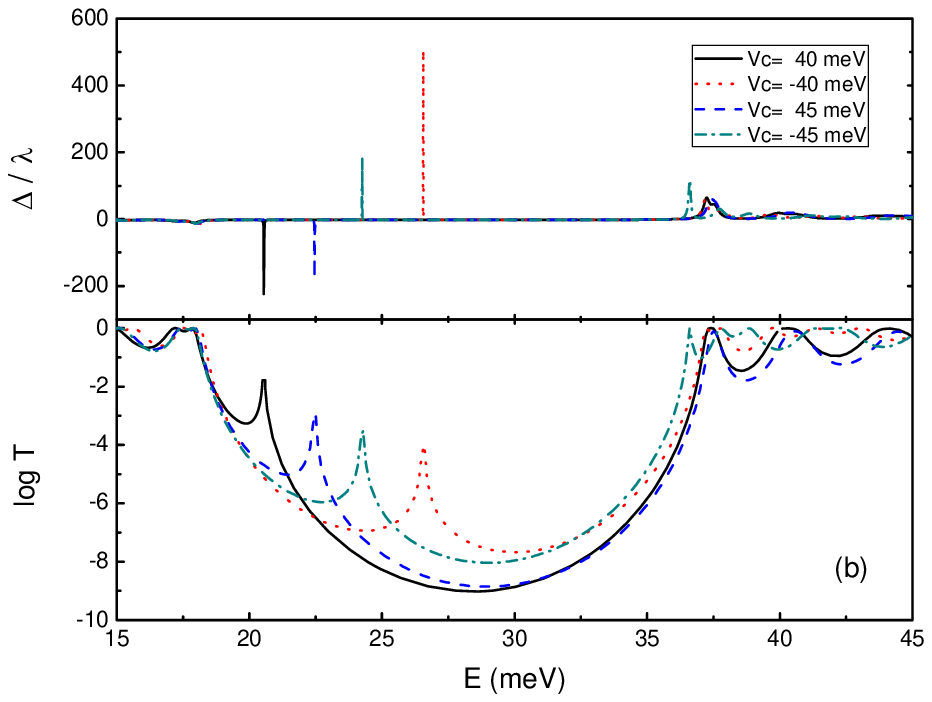}}
\caption{(Color online) Effect of the defect on beam shift and transmission property in graphene superlattice $(AB)^{10}C(BA)^{10}$, where $d_A=d_B=20$ nm, $V_A=50$ meV, $V_B = 0$ meV,  $d_{C}=70$ nm, (a) $V_C = 50$ meV, $\theta_0 =15^\circ$ (solid blue line), $\theta_0 =20^\circ$ (dashed red line)
$\theta_0 =25^\circ$ (dotted black line), (b) $\theta_0 =20^\circ$, $V_C = 40$ meV, (solid black line), $V_C = -40$ meV (dotted red line),
$V_C = 45$ meV (dashed blue line), and $V_C = -45$ meV (dash-dotted green line).}
\label{fig.4}
\end{center}
\end{figure}

Next, we shall investigate the influence of the defect mode on the beam shift. It is found in Fig. \ref{fig.4} (a)
that the beam shift at the incident energy near the defect mode will be greatly enhanced at different incidence angles, and there exists defect mode
with its incident energy inside the zero-$\bar{k}$ bandgap. Actually, the beam shift is almost zero in this band gap, except the near the
corresponding energy of defect mode. Fig. \ref{fig.4} (b) further demonstrates that the enhanced lateral shift can be controlled from negative and
positive by adjusting the potential of defect, $V_C$. It should be noted that the magnitude of the enhancement near the defect mode
can be much larger than that near the edges of band gap. Therefore, the control of beam shifts over the defect provides more
flexibility.

\section{Conclusion}

To summarize, we have investigated the generalized Goos-H\"{a}nchen shifts for the electron beam transmitted through the monolayer graphene superlattice.
The beam shifts near the edges of zero-$\bar{k}$ gap can be negative as well as positive, which depends on the incident energy above or below the Dirac point.
When $q_A d_A = -q_B d_B = m \pi$ ($m=1,2,3...$) is satisfied, there can exist transition between the negative and positive beam shifts.
In addition, the negative or positive beam shifts can be enhanced and controlled by the defect mode when the defect is induced in the graphene superlattice.
These beam shifts in the graphene superlattice are applicable to design various graphene-based electron wave devices.

Last but not least, recent progress on analogous phenomena related to Dirac point in photonic crystals \cite{Wang-Zhu,Beenakker-PRA} or
negative-zero-positive index metamaterial \cite{Chen-PRA} suggest that many exotic phenomena in graphene
can be simulated with relatively simple optical benchtop experiments. Therefore, we also hope that the electronic beam shifts in graphene superlattice presented here can be
tested in the analogy of the photonic crystals containing negative-index and positive-index materials \cite{Wang-Zhu}.


\section*{ACKNOWLEDGEMENT}

We thank A. Lakhtakia for commenting on the manuscript.
This work was supported by the National Natural Science Foundation of China
(Grant Nos. 60806041, 61078021 and 61176118), the Science and Technology Committee of Shanghai Municipality (Grant No. 11ZR1412300),
and the Shanghai Leading Academic Discipline Program
(Grant No. S30105). X. C. also acknowledges Juan de la Cierva
Programme, the Basque Government (Grant No. IT472-10) and Ministerio de
Ciencia e Innovaci\'on (Grant No. FIS2009-12773-C02-01).
L. G. W. acknowledges the National Basic Research Program of China (Grant No. 2012CB921600).


\end{document}